\documentstyle[psfig]{mn0}

\def\simg{\mathrel{\rlap{\raise 0.511ex \hbox{$>$}}{\lower 0.511ex \hbox{$\sim$}}}}
\def\siml{\mathrel{\rlap{\raise 0.511ex \hbox{$<$}}{\lower 0.511ex \hbox{$\sim$}}}}
\def\ep{\varepsilon_p} \def\eg{\varepsilon_\gamma} \def\eG{\varepsilon_{GeV}}
\def\ea{\varepsilon_a} \def\eo{\varepsilon_{\rm o}} \def\e{\varepsilon}
\def\calE{{\cal E}}

\begin{document}

\title [synchrotron self-Compton GRBs] 
   {Prompt GeV emission in the synchrotron self-Compton model for Gamma-Ray Bursts}

\author[A. Panaitescu]{A. Panaitescu \\
       Space Science and Applications, MS D466, Los Alamos National Laboratory,
       Los Alamos, NM 87545, USA}

\maketitle

\begin{abstract}
\begin{small}
 The detection in 10 bursts of an optical counterpart emission (i.e. during the prompt 
 GRB phase) that is $10-10^4$ brighter than the extrapolation of the burst spectrum to 
 optical frequencies suggests a synchrotron self-Compton origin for the GRB emission,
 synchrotron producing the optical counterpart emission. In this model, the second 
 upscattering of the burst photons yields a prompt GeV--TeV emission, whose brightness 
 depends strongly on an unknown quantity, the peak energy of the primary synchrotron 
 spectrum. Measurements of the optical, $\gamma$-ray, and GeV prompt fluxes can be used 
 to test the synchrotron self-Compton model for GRBs and to determine directly the total 
 radiative output of GRBs. For a set of 29 GRBs with optical counterpart detections, 
 we find that the expected GeV photon flux should correlate with the fluence of the 
 sub-MeV emission and anticorrelate with the brightness of the optical counterpart, 
 the strength of these correlations decreasing for an increasing width of the synchrotron 
 peak energy distribution. The detection of a GeV prompt emission consistent with the 
 extrapolation of the burst spectrum to higher energies would rule out the synchrotron 
 self-Compton model if the sub-MeV burst emission were very bright and the (intrinsic) 
 optical counterpart were very dim. 
\end{small}
\end{abstract}
                                                                                   
\begin{keywords}
   radiation mechanisms: non-thermal - shock waves - gamma-rays: bursts
\end{keywords}

\section{Introduction}

 The optical flux during the {\sl prompt} $\gamma$-ray emission is 1--4 orders of magnitude 
larger than the extrapolation of the burst spectrum to optical for GRB 990123 (figure 
2 of Galama et al 1999), GRB 061126 (figure 5 Perley et al 2008), GRB 080319B (figure 
3 of Racusin et al 2008), and for GRBs 060111B, 060927, 061007, 061121, 071003, 080413, 
and 080810 (as can be inferred from the optical and GRB properties listed in Table 1). 
This may suggest that the optical and burst emissions arise from different radiation
processes, synchrotron emission dominating the optical {\sl counterpart} and inverse-Compton
scatterings producing the 10 keV--10 MeV emission (model b2 of M\'esz\'aros \& Rees 1997,
Panaitescu \& M\'esz\'aros 2000).

 An essential {\sl assumption} for the synchrotron self-Compton interpretation of GRBs is 
that the optical counterpart and burst emissions arise from the same relativistic ejecta.
For GRB 080319B, whose optical counterpart emission was well-sampled (Karpov et al 2008), 
that assumption is supported by the broad correlation of optical and $\gamma$-ray prompt 
light-curves (Stamatikos et al 2008). A correlation between burst and optical counterpart 
emissions is also possible in the internal shock model (Rees \& M\'esz\'aros 1994) for 
GRBs if the pair of reverse and forward shocks produced by the interaction of relativistic 
shells radiate in the optical and at sub-MeV, as was proposed by Yu, Wand \& Dai (2008) 
for GRB 080319B. The latter model requires that, for all pairs of interacting shells, 
the Lorentz factor ratio is very larger (above 1000), but it produces a weaker GeV emission 
from inverse-Compton scatterings than does the second upscattering of the former model.
A tight correlation of GRB and optical counterpart fluctuations is not expected in either
model, as the spectra of two emission components (synchrotron and inverse-Compton, or just 
synchrotron from reverse and forward shocks, respectively) may peak, sometimes, far from 
the corresponding observing band-passes (optical and $\gamma$-ray) and not yield a pulse 
in that photon range.

 From the optical and $\gamma$-ray properties of the prompt emissions of GRB 080319B,
Kumar \& Panaitescu (2008) have inferred that the upscattering of GRB photons (i.e.
the second inverse-Compton scattering of the primary synchrotron photons) should have
produced a GeV photon yield over the burst duration of thousands of photons for 
Fermi's Large Area Telescope (LAT) and hundreds of photons for Agile's Gamma-Ray
Imaging Detector (GRID), the second scattering GeV--TeV emission accompanying GRB 
080319B containing 10 times more energy than released at sub-MeV by the first scattering. 
If the synchrotron self-Compton process were at work in other bursts with an optical 
counterpart dimmer than that of GRB 080319B, then the Compton parameter for the second 
scattering could be substantially larger than for GRB 080319B, leading to bursts that
radiate much more energy in the GeV than at sub-MeV (Piran, Sari \& Zou 2008); however, 
synchrotron peak energies well from optical can reduce substantially the Compton parameter 
of the second scattering and its GeV flux.

 Currently, the observational evidence for a prompt emission component peaking above 
10 MeV (as is possible in the synchrotron self-Compton model for GRBs) is modest. 
The spectra of 15 GRBs measured by the Energetic Gamma-Ray Experiment Telescope (EGRET) 
calorimeter on the Compton Gamma-Ray Observatory up to 100 MeV (Kaneko et al 2008) 
show only 3 such cases. One of them is GRB 941017 (Gonzales et al 2003), whose 
$\nu F_\nu$ spectrum rises up to 100 MeV; the other ones are GRB 930506 and 980923. 
A prompt GeV flux that exceeds the extrapolation of the burst spectrum to higher 
energies has also been detected by EGRET for GRB 940217 (Hurley et al 1994). 
At the other end, the most notable evidence provided by EGRET for the absence of
higher energy emission is for GRB 930131 (Sommer et al 1994), whose power-law 
spectrum extends up to 1 GeV. We also note that the prompt emissions above 100 MeV 
of two recent bursts measured by Fermi-LAT lie on the extrapolation of the MeV spectrum. 

 Double upscattering of the synchrotron emission is not the only model that can
yield a prompt GeV emission. Previous proposed models for a {\sl prompt} GeV emission 
include the more "mundane" synchrotron and inverse-Compton from internal shocks 
(e.g. Papathanassiou \& M\'esz\'aros 1996), inverse-Compton emission from the 
reverse-shock (e.g. Granot \& Gueta 2003) or from the forward-shock (e.g. Wang, Dai 
\& Lu 2001), and upscattering of reverse-shock synchrotron photons in the forward-shock 
(Pe'er \& Waxman 2004), as well as some more "exotic" and uncertain ones 
(e.g. synchrotron emission from ultra-high energy protons or the electrons and muons 
formed from by the photo-pion decay of those protons -- Asano \& Inoue 2007).
 Evidently, a comparison of the optical, sub-MeV, and GeV emissions with model-expected 
correlations will be required to distinguish among the various process proposed for the 
higher energy component.

 In this paper, we develop the formalism by which optical counterpart and prompt burst 
measurements can be used to infer the GeV flux accompanying GRBs and apply it to the 
bursts with optical counterpart measurements (detections or upper limits) to calculate 
the bolometric GRB output. As shown below, these quantities depend strongly on the peak 
energy of the primary synchrotron spectrum. The direct determination of that quantity 
through optical and near-infrared observations of the prompt emission and the measurement 
of the GeV prompt flux can then be used to test the synchrotron self-Compton model for 
GRBs. If the peak energy of the synchrotron spectrum cannot be determined observationally, 
then the GeV and optical fluxes and spectral slopes can be used to perform a weaker test 
of that model.

 The following calculations for the synchrotron self-Compton emissions are general
and do not depend on the dissipation mechanism (i.e. type of shock) which accelerates 
relativistic electrons and produces magnetic fields. It could be the external reverse 
shock which propagates into the relativistic ejecta, if that mechanism can account for 
the burst variability, or it could be internal shocks in a variable outflow, as was 
proposed by Sari \& Piran (1999) and M\'esz\'aros \& Rees (1999), respectively, to
explain the bright optical counterpart of GRB 990123. The physical parameters of the
synchrotron self-Compton model required to account for the optical and sub-MeV emissions 
of that particular burst, GRB 990123, were inferred by Panaitescu \& Kumar (2007).
As for GRB 080319B, it was found that the peak energy of the synchrotron spectrum
was not far from the optical.

\section{Formalism}

 In the synchrotron self-Compton model for the GRB emission, the peak energy and peak
flux of the first inverse-Compton scattering are the peak energy $\eg$ and flux $F_\gamma$
of the GRB spectrum. The peak energy $\ep$ and flux $F_p$ of the primary synchrotron 
spectrum could be measured directly with robotic telescopes performing multiband observations
of the optical counterpart only if $\ep$ falls in the optical bandpass (i.e. $\ep \sim
2$ eV), but otherwise remain unknown (optical counterpart measurements yield a relation 
between $F_p$ and $\ep$). Both quantities $F_p$ and $\ep$ are needed to calculate the 
typical energy $\gamma_p m_e c^2$ of the radiating electrons and the optical thickness 
$\tau_e$ to electron scattering of the radiating medium, which, together with the 
$F_\gamma$ and $\eg$ of the first scattering, lead to the peak energy $\eG$ and flux 
$F_{GeV}$ of the second inverse-Compton scattering. The last two quantities set the GeV 
prompt flux, thus observations by Fermi-LAT and Agile-GRID of the GeV emission accompanying 
GRBs can be used in conjunction with the optical counterpart and burst measurements to 
test the synchrotron self-Compton model for GRBs. In this section, we relate the properties 
of the twice upscattered emission to those of the prompt optical and $\gamma$-rays. 

 The peak energy $\eG$ of the second inverse-Compton emission spectrum and the peak flux 
$F_{GeV}$ at $\eG$ are related to those of the first inverse-Compton by 
\begin{equation}
 \eG = \gamma_p^2 \eg \; \quad F_{GeV} = \tau_e F_\gamma 
\label{Gev}
\end{equation}
with $\gamma_p = (\eg/\varepsilon_{peak})^{1/2}$ 
and $\tau_e = F_\gamma/F_{peak}$ relating the peak energies and flux of the first 
inverse-Compton spectrum to those of the spectrum of the {\sl received} synchrotron 
emission, $\varepsilon_{peak}$ and $F_{peak}$. If the emitting fluid is optically thin
to synchrotron self-absorption at the peak energy $\ep$ of the synchrotron {\sl emissivity},
then $\varepsilon_{peak}=\ep$ and $F_{peak}=F_p$; however, if the optical thickness
to synchrotron self-absorption at $\ep$ is above unity, the received spectrum peaks
at the synchrotron self-absorption energy $\ea$ (i.e. $\varepsilon_{peak} = \ea$).
Thus
\begin{equation}
 \gamma_p =  \left\{ \begin{array}{ll} 
             \hspace*{-2mm}  (\eg/\ep)^{1/2} & \tau_p < 1 \\ 
             \hspace*{-2mm}  (\eg/\ea)^{1/2} & \tau_p > 1 
             \end{array} \right.
   \; , \quad
 \tau_e = \left\{ \begin{array}{ll}
             \hspace*{-2mm} F_\gamma/F_p & \tau_p < 1 \\ 
             \hspace*{-2mm} F_\gamma/F_a & \tau_p > 1
             \end{array} \right.
\label{gpte}
\end{equation}
where $F_a$ is the synchrotron flux at $\ea$ and
\begin{equation}
 \tau_p  = \frac{5e \tau_e}{\sigma_e B \gamma_p^5}
\end{equation}
is the optical thickness to synchrotron self-absorption at the peak energy $\ep$,
$\sigma_e$ being the cross-section for electron scattering (in the Thomson regime)
and $B$ the magnetic field strength. The value of $B$ can be inferred from the 
synchrotron peak energy:
\begin{equation}
 \ep = \frac{eh}{4m_ec} \frac{\gamma_p^2 B \Gamma}{z+1}
\end{equation}
taking into account the relativistic boost of photon energy by the Lorentz factor
$\Gamma$ of the source, which leads to
\begin{equation}
 \tau_p = \frac{52.6\; {\rm MeV}}{\ep} \frac{\Gamma\tau_e}{(z+1)\gamma_p^3} \;.
\label{taup}
\end{equation}

 Therefore, to find $\gamma_p$ and $\tau_e$, the quantities $\ep$, $\ea$, $F_p$, and $F_a$
must be constrained from the counterpart optical flux (which is the only measurable
quantity directly pertaining to the synchrotron emission). As that provides only one
constraint, we shall express the following results as function of the peak energy $\ep$,
and consider separately the $\tau_p < 1$ and $\tau_p >1$ cases.

\begin{figure}
\centerline{\psfig{figure=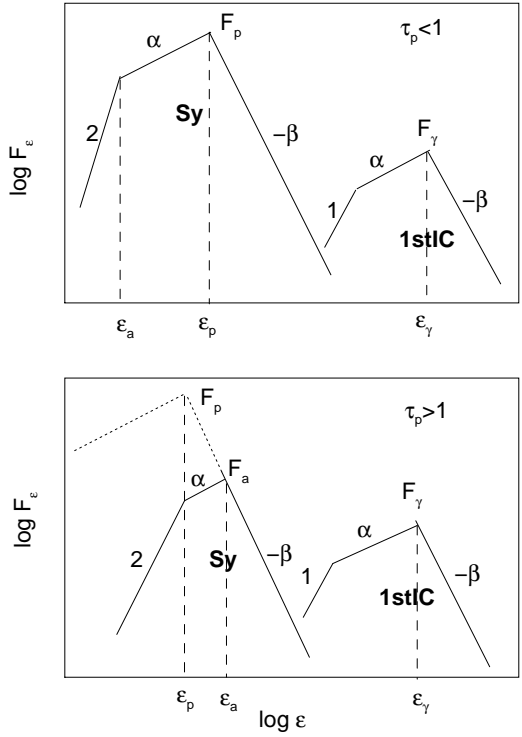,width=7cm}}
\caption{ Power-law spectra of synchrotron and first inverse-Compton emissions for
    an emitting plasma which is optically thin ($\tau_p <1$, upper panel) and optically 
    thick ($\tau_p >1$, lower panel) at the emissivity peak energy $\ep$, and for 
    electrons with a power-law energy distribution above that corresponding to the 
    synchrotron characteristic energy $\ep$. Spectral power-law indices ($d\log F_\e/
    d\log \e$) are indicated. 
    Below the peak of the spectrum, the slopes are $\alpha=1/3$ and $\alpha=5/2$ 
    for $\tau_p <1$ and $\tau_p >1$, respectively. However, we allow an arbitrary
    slope $-\beta < \alpha < 1/3(5/2)$ to accommodate the diversity of low-energy slope
    measured for the GRB emission (which is the first inverse-Compton component), which
    may be due to a more complex electron distribution than a pure power-law only above 
    $\gamma_p$. GRB observations determine the peak energy and flux ($\eg$, $F_\gamma$)
    of the first inverse-Compton emission, while optical counterpart measurements set a
    constraint on the spectral peak properties ($\ep$, $F_p$) of the synchrotron
    emission which depends on the location of the optical bandpass relative to the
    break energies $\ep$ (emissivity peak) and $\ea$ (self-absorption).} 
\label{f1}
\end{figure}

 For $\tau_p < 1$, equations (\ref{gpte}) and (\ref{taup}) yield
\begin{equation}
 \tau_p = k \frac{F_\gamma}{F_p} \ep^{1/2}\;, \quad 
          k \equiv \frac {52.6\, \Gamma\; {\rm MeV}}{(z+1) \eg^{3/2}} \;.
\label{tp1}
\end{equation}
In this case, the self-absorption energy is 
\begin{equation}
 \ea = \ep \tau_p^{3/5} < \ep 
\label{ea1}
\end{equation}
and the optical counterpart flux $F_o$ is (Figure \ref{f1}, upper panel)
\begin{equation}
 F_o =  F_p \left\{ \begin{array}{ll}  
         (\eo/\ea)^2 (\ea/\ep)^\alpha & \eo < \ea < \ep    \\
         (\eo/\ep)^\alpha             & \ea < \eo < \ep    \\
         (\ep/\eo)^\beta              & \ea < \ep < \eo    
        \end{array} \right.
\label{Fo1}
\end{equation} 
where $\alpha$ and $\beta$ are the spectral slopes below and above $\ep$ of the
synchrotron emissivity, which are the same as low and high energies slope of the 
first inverse-Compton GRB spectrum: $F_\varepsilon \propto \varepsilon^\alpha$ and 
$F_\varepsilon \propto \varepsilon^{-\beta}$, respectively. 
Substituting $F_p$ in equation (\ref{tp1}) and using equation (\ref{ea1}), one finds 
\begin{equation}
 \tau_p = \left( k \frac{F_\gamma}{F_o} \frac{\eo^2}{\ep^{3/2}} \right)^{5/(11-3\alpha)} \;.
\label{tp1}
\end{equation}
for the $\eo < \ea < \ep$ case. Then, the starting condition $\tau_p <1$ is 
equivalent to $\ep > \e_1$ with 
\begin{equation}
 \e_1 \equiv (k' \eo^{-2} )^{-2/3} \;, \quad k' \equiv \frac{F_o}{k F_\gamma} 
\end{equation}
while the assumption $\eo < \ea$ is equivalent to $\ep > \e_2$ with
\begin{equation}
 \e_2 \equiv (k' \eo^{\frac{5}{3}-\alpha})^{6/(13-6\alpha)} \;.
\end{equation}
For the $\ea < \eo < \ep$ case, one obtains
\begin{equation}
 \tau_p =  k \frac{F_\gamma}{F_o} \frac{\eo^\alpha}{\ep^{\alpha-\frac{1}{2}}} 
\label{tp2}
\end{equation}
while for the $\ea < \ep < \eo$ case
\begin{equation}
 \tau_p = k \frac{F_\gamma}{F_o} \frac{\ep^{\beta+\frac{1}{2}}}{\eo^\beta} 
\label{tp3}
\end{equation}
the $\tau_p <1$ condition requiring that $\ep < \e_3$, where
\begin{equation}
 \e_3 \equiv (k' \eo^\beta)^{2/(2\beta+1)} \;.
\end{equation}
The three reference photon energies $\e_1$, $\e_2$, and $\e_3$ depend only on observables
and allow the selection of the photon energy ordering given in equation (\ref{Fo1}):
\begin{equation}
 \left\{ \begin{array}{lll}
        \eo < \ea < \ep  & {\rm if} & \e_1, \e_2 < \ep  \\
        \ea < \eo < \ep  & {\rm if} & \eo < \ep < \e_1  \\
        \ea < \ep < \eo  & {\rm if} & \ep < \e_3, \eo
        \end{array} \right.
\label{case1}
\end{equation}
Thus, given a peak energy $\ep$, equation (\ref{case1}) identifies the ordering
of $\eo$, $\ea$, and $\ep$, from where $\tau_p$ can be determined using equations 
(\ref{tp1}), (\ref{tp2}), or (\ref{tp3}), further leading to $\ea$ through equation
(\ref{ea1}), then to the peak flux $F_p$ of equation (\ref{Fo1}), then to the $\tau_e$ of
equation (\ref{gpte}) and, finally, to the spectral peak flux of the second inverse-Compton
scattering of equation (\ref{Gev}) as a function of $\ep$.
 
 For $\tau_p >1$, we are interested in calculating the self-absorption energy $\ea$
and the synchrotron flux $F_a$ at $\ea$ as a function of $\ep$. From equations
(\ref{gpte}) and (\ref{taup}), one obtains 
\begin{equation}
 \tau_p = k \frac{F_\gamma}{F_a} \frac{\ea^{3/2}}{\ep}
\label{tp2}
\end{equation}
with $\ea$ being
\begin{equation}
 \ea = \ep \tau_p^{\frac{2}{2\beta+5}} > \ep \;.
\label{ea2}
\end{equation}
The optical counterpart flux is related to $F_a$ through (Figure \ref{f1}, lower panel)
\begin{equation}
 F_o =  F_a \left\{ \begin{array}{ll}
         (\eo/\ep)^2 (\ep/\ea)^\alpha & \eo < \ep < \ea    \\
         (\eo/\ea)^\alpha             & \ep < \eo < \ea    \\
         (\ea/\eo)^\beta              & \ep < \ea < \eo
        \end{array} \right.
\label{Fo2}
\end{equation}
Continuing in a similar way as shown above for $\tau_p <1$, one finds that the ordering
of energies in equation (\ref{Fo2}) is set by $\ep$ as following 
\begin{equation}
 \left\{ \begin{array}{lll}
        \eo < \ep < \ea  & {\rm if} & \eo <\ep < \e_2   \\
        \ep < \eo < \ea  & {\rm if} & \e_4 < \ep < \eo  \\
        \ep < \ea < \eo  & {\rm if} & \e_3 < \ep < \e_4
        \end{array} \right.
\label{case2}
\end{equation}
where
\begin{equation}
 \e_4 \equiv (k' \eo^{\beta+1})^{2/(2\beta+3)} \;.
\end{equation}
Then, the optical thickness to self-absorption at the spectral peak of the synchrotron
emissivity is
\begin{equation}
 \tau_p = \left\{ \begin{array}{lll}
    (k \eo^2/\ep^{3/2})^{(\beta+5/2)/(\beta+\alpha+1)} &  \eo < \ep < \ea \\
    (k \eo^\alpha/\ep^{\alpha-1/2})^{(\beta+5/2)/(\beta+\alpha+1)} & \ep < \eo < \ea \\
    (k \ep^{\beta+1/2}/\eo^\beta)^{\beta+5/2} & \ep < \ea < \eo \\
       \end{array} \right.
\label{taup1}
\end{equation}
Equations (\ref{taup1}), (\ref{ea2}), and (\ref{Fo2}) allow the calculation of $\ea (\ep)$
and $F_a(\ep)$. 

 The reference energies $\e_1$, $\e_2$, $\e_3$, and $\e_4$ have the same form
$\e(x) = (k' \eo^x)^{2/(2x+1)}$ with $x=-2,\frac{5}{3} -\alpha,\beta,\beta+1$,
respectively, thus $\e(x)$ has a singularity at $x=-1/2$. For $k' < \eo^{1/2}$, 
i.e. $F_o/F_\gamma < (z+1) \eg^{3/2}/(52.6\, \Gamma \eo^{1/2}\, {\rm MeV})$, it can be 
shown that $d\e(x)/dx > 0$ and the reference energies ordering is $\e_3 < \e_4 < \eo < \e_1$
(the relative location of $\e_2$ depending on $\alpha$). For $k' > \eo^{1/2}$,
$d\e(x)/dx < 0$ and $\e_1 < \eo < \e_4 < \e_3$.

 The Klein-Nishina effect on the second inverse-Compton scattering is important if the
energy of the first scattering (GRB) photon, as measured in the electron frame, is 
comparable or larger than the electron rest-mass energy, i.e. if $(z+1)(\eg/\Gamma) 
\gamma_p > m_e c^2$, where the typical electron Lorentz factor $\gamma_p$ is obtained
using equation (\ref{gpte}). Considering only the $\tau_p < 1$ case, for which $\gamma_p 
= (\eg/\ep)^{1/2}$, implies that the Klein-Nishina effect is important for 
$\ep < \e_{kn}$ with
\begin{equation}
 \e_{kn} \equiv \left( \frac{z+1}{\Gamma m_e c^2}\right)^{\hspace*{-1mm}2} 
          \hspace*{-1mm}\eg^3 =
      3 \left(\frac{\eg}{200\,{\rm keV}}\right)^{\hspace*{-1mm}3} 
     \hspace*{-1mm} \left(\frac{z+1}{3}\right)^{\hspace*{-1mm}2}  
     \hspace*{-1mm} \left(\frac{\Gamma}{300}\right)^{\hspace*{-1mm}-2} 
     \hspace*{-1mm} {\rm eV} \;.
\end{equation}

 Thus, the Klein-Nishina effect is expected to be important only if the peak energy
of the synchrotron spectrum is below optical. 
 In this case, the energy of the twice upscattered photon is $\eG = \Gamma m_e c^2 
\gamma_p/(z+1)$ (lower than given in equation \ref{Gev}) and the peak flux of the 
second inverse-Compton emission is diminished by the decreased scattering
cross-section, $\tau_{e,kn} \simeq \tau_e (\ep/\e_{kn})^{1/2} < \tau_e$. 
For $\ea < \ep < \eo$, equations (\ref{Fo1}) and (\ref{gpte}) lead to
$\tau_e = (F_\gamma/F_o) (\ep/\eo)^\beta$, and the Compton parameter for the second 
scattering is $Y_{GeV} = \tau_{e,kn} \eG/\eg = \tau_e (\eg/\e_{kn})^2$.

 For $\ep$ above optical, the Klein-Nishina effect is negligible and $Y_{GeV} = Y_\gamma = 
(F_\gamma \eg)/ (F_p \e_p)$, where $Y_\gamma$ is the Compton parameter for the first
scattering and $F_p$ is given by equation (\ref{Fo1}) for $\ea < \eo < \ep$.
Thus, the fluence of the twice upscattered emission, $\Phi_{GeV} = Y_{GeV} \Phi_\gamma$, 
is
\begin{equation}
 \Phi_{GeV} = \frac{\Phi_\gamma^2}{t_\gamma F_o} \times  \left\{ \begin{array}{ll}
  \hspace*{-2mm} \left(\frac{\displaystyle \Gamma m_e c^2}{\displaystyle z+1}\right)^2 
        \frac{\displaystyle \ep^\beta}{\displaystyle \eg^3 \eo^\beta} & \ep < \e_{kn} \\ 
  \hspace*{-2mm} \frac{\displaystyle \eo^\alpha}{\displaystyle \ep^{\alpha+1}} & \e_{kn} < \ep 
   \end{array} \right.
\label{PhiGeV}
\end{equation}
where $\Phi_\gamma$ and $t_\gamma$ are the GRB fluence and duration, respectively.

 Therefore, for fixed properties of the prompt optical and GRB emissions, the fluence of 
the GeV emission and its Compton parameter increase as $\ep^\beta$ for $\ep < \e_{kn}$ 
and decrease as $\ep^{-(\alpha+1)}$ for $\ep > \e_{kn}$, being maximal when the peak 
energy of the synchrotron spectrum is in or close to the optical 
($\ep = \e_{kn} \simeq \eo$).
Using equation (\ref{PhiGeV}) to assess the effect of observables on the expected
fluence of the second scattering, it could be expected that the GeV fluence is
(1) {\sl correlated} with the burst fluence (which is quite trivial, as the GeV photons 
are the upscattered burst photons) and 
(2) {\sl anticorrelated} with the optical counterpart flux, burst duration, and
peak energy of the GRB spectrum (if $\ep$ is below optical), 
with the caveats that these correlations 
(a) should be weakened and could be even wiped out by variations in $\ep$ from burst 
to burst (as $\ep$ affects strongly the GeV flux),
(b) could be affected by correlations among optical and burst properties. 

\begin{figure}
\psfig{figure=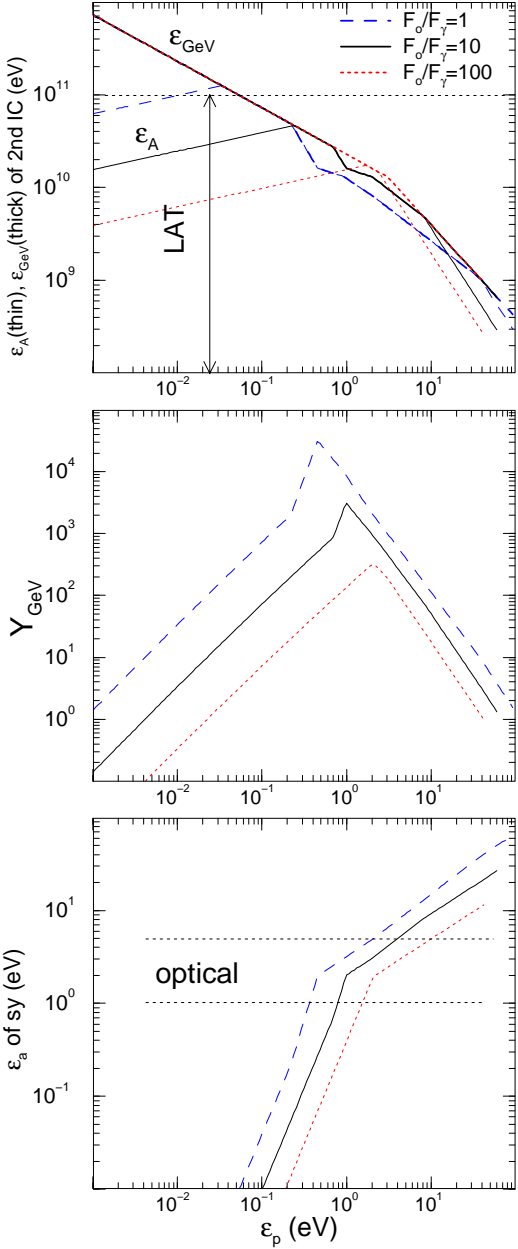,width=7cm}
\caption{Dependence of twice up-scattered self-absorption $\e_A$ and peak photon energy
   $\eG$ (top panel), Compton parameter for second scattering $Y_{GeV}$ (mid panel), and 
   synchrotron self-absorption energy $\ea$ (bottom panel) on the peak energy $\ep$ of 
   the synchrotron emissivity, for three likely ratios of the (synchrotron) optical 
   counterpart flux $F_o$ to the (first inverse-Compton) GRB flux $F_\gamma$ at the peak 
   of GRB spectrum. 
   Other parameters are set to average values for the GRBs of Table 1 for which optical
   counterpart measurements have been obtained: $F_\gamma = 0.3$ mJy, GRB peak energy 
   $\eg = 200$ keV (low and high-energy GRB spectral slopes $\alpha=0$ and $\beta = 1.5$
   were assumed), for calculation of Klein-Nishina effect on the upscattered emission: 
   redshift $z=2$ and a source Lorentz factor $\Gamma=300$ was assumed.
   The $\ep$ is upper limited by the condition that the 10 keV (lower bound of bandpass 
   of most burst detectors) flux is dominated by the first inverse-Compton component. 
   }
\label{f2}
\end{figure}

 Figure \ref{f2} shows the dependence of some characteristics of the second inverse-Compton
scattering (peak energy $\eG = \gamma_p^2 \eg$ and upscattered self-absorption frequency 
$\e_A = \gamma_p^4 \ea$ in the upper panel, Compton parameter $Y_{GeV} = (\eG F_{GeV})/
(\eg F_\gamma)$ in the mid panel) on the peak energy $\ep$ of the synchrotron spectrum, 
calculated with the aid of the equations above.
Aside from the unknown $\ep$, the optical and GRB spectral peak fluxes, $F_o$ and $F_\gamma$,
are the other major factors affecting the brightness of the twice upscattered emission,
the other parameters and observables having a lesser effect. 

 The GeV emission spectrum has the same shape as given in equations (\ref{Fo1}) and
(\ref{Fo2}), except that upscattering of the self-absorbed part of the synchrotron
spectrum yields a flatter one, $F_\e \propto \e$ (Panaitescu \& M\'esz\'aros 2000). 
 Middle panel of Figure \ref{f2} shows the expected behaviour of the second scattering's
Compton parameter, peaking for $\ep \sim 1$ eV. As $Y_{GeV}$ is a measure of the GeV 
fluence, it follows that a measurement of the prompt GeV fluence yields two solutions 
for the unknown peak $\ep$ of the synchrotron spectrum (see also equation \ref{PhiGeV}). 
The real solution can be find using the 1--100 GeV spectrum: a hard $F_\epsilon \propto 
\epsilon$ spectrum (resulting from upscattering twice synchrotron photons below 
self-absorption) or one of slope $\alpha$ up to tens of GeV indicates that $\ep \siml 1$ 
eV, while and a soft spectrum of slope $\beta$ above 1--10 GeV is expected for $\ep 
\simg 10$ eV. Then, compatibility with the spectrum of the optical counterpart emission,
which the bottom panel of Figure \ref{f2}) shows that should be optically thin for 
$\ep \siml 1$ eV and thick for $\ep \simg 1$ eV, 
offers a possible test of the synchrotron self-Compton model for the burst emission. 
A stronger test can be done if the peak energy of the GeV spectrum is measured, as in 
this case the GeV peak energy and fluences provide two independent constraints on 
the peak energy of the synchrotron spectrum.

\section{Application to bursts with optical counterpart measurements}

 Without knowing the peak energy of the synchrotron spectrum, we proceed to estimate 
the GeV output and total energetics of bursts for which optical counterpart measurements 
exist for $\ep = 1$ eV, which maximizes the GeV prompt output, and for $\ep = 0.01$ eV
(values above optical also reduce the GeV output, with an upper limit of 100 eV on
$\ep$ is imposed by requiring that the 10 keV prompt emission is dominated by the first
scattering and not by the high-energy tail of the synchrotron spectrum).

 The relevant properties of the optical and GRB prompt emissions of bursts with optical
counterpart measurements are listed in Table 1. For more than half of those bursts,
only upper limits on the optical counterpart flux have been obtained, the upper limit
listed in Table 1 one being the deepest available and over an integration time that
is within the burst emission or up to a factor two in time after the last GRB peak.
Optical counterpart measurements have been corrected for the often modest Galactic
dust extinction (given in last column), but may be affected by a more substantial
extinction in the host galaxy that could be estimated, in some cases, from the optical 
afterglow spectrum. 

\begin{table*}
\vspace*{-3mm}
\caption{Gamma-ray (columns 3--7) and optical (counterpart in column 8, afterglow in
  columns 9 and 10) properties of GRBs with optical counterpart measurements.
  $t_\gamma$ = burst duration, $\Phi_\gamma$ = burst fluence, $\eg$ = GRB peak energy,
  $\alpha$ = slope of GRB spectrum below $\eg$ ($F_\e \propto \e^\alpha$).
  For GRBs in boldface, the optical counterpart flux is more than 10 times larger than 
  the extrapolation of the GRB spectrum to optical energies.  
  "Wh" is for UVOT's white filter, "Un" for unfiltered. }
\vspace*{-2mm}
\begin{tabular}{ccccccccccccccccc}
  \hline 
 GRB   & redshift & $t_\gamma$ & $\Phi_\gamma$ & band & $\alpha$ & $\eg$ & OC flux &  
      $t_b$ & decay & E(B-V) \\
       &     &   (s)    &   (cgs) & (keV)&      &  (keV)   &  (mag)  &  (d) &  index &    \\
  \hline
{\bf080810}&3.35&140& 1.7e-5 &20-1000&-0.2& 550 & Un=13.2 &$>$5.6 &  1.40 & 0.03 \\
080802 &    &176& 1.3e-6 &15-150 &-0.8&     & Wh$>$20 &       &       & 0.80 \\
080607 &3.04&85 & 8.9e-5 &20-4000&-0.1& 420 & R=15.2  &$>$0.6 &       & 0.02 \\
080603B&2.69&70 & 4.5e-6 &20-1000&-0.2& 100 & Un=14.1 &  0.6  &  3.05 & 0.01 \\
{\bf080413}&2.44&55 & 4.8e-6 &15-1000&-0.2& 170 & Un=12.8 &$>$0.15&  1.2  & 0.16 \\
{\bf080319B}&0.94&57 & 5.7e-4 &20-7000& 0.2& 650 & V=6.0   &$>$20  &  1.33 & 0.01 \\
080310 &2.43&365& 2.3e-6 &15-150 &    &$<$30& R=17    &  2    &  2.4  & 0.04 \\
080307 &    &64 & 7.3e-7 &15-150 &-0.4&     & R$>$16.9&$>$0.06&  0.7  & 0.03 \\
080229 &    &64 & 9.0e-6 &15-150 &-0.9&     & R$>$14.7&       &       & 0.15 \\
080212 &    &123& 2.9e-6 &15-150 &-0.6&     & R$>$17.8&$>$0.6 &  0.4  & 0.16 \\
080205 &    &107& 2.1e-6 &15-150 &-1.1&     & Un=18.1 &       &       & 0.09 \\
071031 &2.69&180& 9.0e-7 &15-150 &    &$<$30& R=15    &$>$0.13&  0.55 & 0.01 \\
071025 &    &109& 6.5e-6 &15-150 &-0.8&     & R$>$17.3&$>$0.2 &  1.8  & 0.08 \\
071011 &    &61 & 2.2e-6 &15-150 &-0.4&     & R$>$16.9&$>$0.15&  0.7  & 0.91 \\
{\bf071003}&1.60&30 & 1.2e-5 &20-4000& 0.1&  800& Un=12.8 &$>$7.9 &  1.60 & 0.15 \\
070808 &    &32 & 1.2e-6 &15-150 &-0.5&     &Un$>$16.2&       &       & 0.02 \\
070721B&3.63&32 & 2.1e-6 &15-150 &-0.3&     & Wh=15.9 &       &       & 0.02 \\
070621 &    &40 & 4.3e-6 &15-150 &-0.6&     &Un$>$16.6&       &       & 0.05 \\
070616 &    &402& 1.9e-5 &15-150 &-0.9&  100& V=16.5  &       &       & 0.40 \\
070521 &0.55&55 & 1.8e-5 &20-1000& 0.1&  220& R$>$17.1&       &       & 0.03 \\
070429 &    &163& 9.2e-7 &15-150 &-1.1&     &Un$>$16.2&       &       & 0.17 \\
070420 &    &120& 2.6e-5 &20-1000&-0.1&  170& R=16.2  &$>$0.15& 0.88  & 0.52 \\
070419B&    &91 & 1.1e-5&100-1000& 0.1&     &Wh$>$18.5&       &       & 0.09 \\
070419A&0.97&116& 5.6e-7 &15-150 &    &$<$30& R$>$18.6&$>$3.7 & 0.99  & 0.03 \\
070411 &2.95&101& 2.5e-6 &15-150 &-0.7&     & R=17.9  &$>$5.8 & 1.11  & 0.29 \\
070306 &1.50&210& 5.5e-6 &15-150 &-0.7&     &Wh$>$19.8&       &       & 0.03 \\
070220 &    &30 & 1.1e-5 &20-2000&-0.2&  300&Wh$>$19.6&       &       & 0.90 \\
070208 &1.17&48 & 4.3e-7 &15-150 &-1.0&     &Un$>$18.7&$>$0.3 & 0.55  & 0.01 \\
070129 &    &460& 3.1e-6 &15-150 &-1.0&     & V$>$17.3&       &       & 0.14 \\
061222 &    &100& 2.7e-5 &20-2000& 0.1&  280&Un$>$17.0&       &       & 0.10 \\
{\bf061126}&1.16&25 & 2.0e-5 &30-2000& 0.1&  935& R=12.93 &$>$1.8 & 0.99  & 0.18 \\
{\bf061121}&1.31&81 & 1.4e-5 &15-150 & 0.2&  455& Un=14.9 &$>$3.9 & 1.05  & 0.05 \\
061110 &0.76&41 & 1.1e-6 &15-150 &-0.7&     &Un$>$16.2&       &       & 0.09 \\
{\bf061007}&1.26&90 & 2.5e-4&20-10000& 0.3&  400& Un=13.6 &$>$1.7 & 1.70  & 0.02 \\
{\bf060927}&5.47&23 & 1.1e-6 &15-150 & 0.1&   70& Un=16.5 &$>$2.6 & 1.01  & 0.06 \\
060904B&0.70&192& 1.7e-6 &15-150 &-0.7&     & Un=17.3 &$>$1.9 & 1.02  & 0.17 \\
060904A&    &80 & 1.6e-5 &10-2000& 0.1&  160& R$>$16.5&       &       & 0.02 \\
060814 &0.84&134& 2.7e-5 &20-1000&-0.4&  260&Wh$>$19.7&       &       & 0.04 \\
060729 &0.54&116& 2.7e-6 &15-150 &-0.9&     & Un=15.67& $>$28 & 1.27  & 0.05 \\
060719 &    &55 & 1.6e-6 &15-150 &-1.0&     & z$>$16.6&       &       & 0.07 \\
060714 &2.71&115& 3.0e-6 &15-150 &-1.0&     & Wh=19.2 &$>$3.3 & 1.22  & 0.08 \\
060607 &3.08&100& 2.6e-6 &15-150 &-0.5&     & r=16.3  &$>$0.3 & 1.20  & 0.03 \\
060602 &0.79&60 & 1.6e-6 &15-150 &-0.1&     & R$>$15  &       &       & 0.03 \\
060507 &    &185& 4.1e-6 &15-150 &-0.8&     &Un$>$15.5&       &       & 0.16 \\
060418 &1.49&44 & 1.6e-5 &20-1100&-0.5&  230& z=15.3  &$>$1.2 & 1.25  & 0.22 \\
060312 &    &43 & 1.8e-6 &15-150 &-0.4&     & R$>$14.6&       &       & 0.19 \\
060210 &3.91&255& 7.7e-6 &15-150 &-0.5&     & R$>$17.5&       &       & 0.09 \\
060124 &2.30&710& 2.8e-5 &20-2000&-0.3&  335& V=17.08 &$>$6.2 & 1.42  & 0.14 \\
{\bf060111B}&    &25 & 5.6e-8&100-1000&-0.5&     & R=13.8  &       &       & 0.10 \\
051117 &    &140& 4.6e-6 &15-150 &-0.8&     & V=20.0  &       &       & 0.02 \\
051111 &1.55&31 & 8.4e-6&100-700 &-0.5&     & R=13.2  &$>$1.0 & 1.62  & 0.16 \\
051022 &0.80&200& 2.6e-4 &20-2000&-0.2&  510& R$>$17.4&       &       & 0.07 \\
051001 &    &190& 1.8e-6 &15-150 &-1.1&     & R$>$16.2&       &       & 0.02 \\
050915 &    &53 & 8.8e-7 &15-150 &-0.4&     & R$>$17.4&       &       & 0.03 \\
050904 &6.29&225& 5.4e-6 &15-150 &-0.4&  340& R=18.5  &$>$5.3 & 1.15  & 0.06 \\
050822 &    &102& 3.4e-6 &15-350 &    &$<$30& R$>$16.6&       &       & 0.02 \\
050714 &    &40 & 6.2e-7 &20-200 &    &     & R$>$16.6&       &       & 2.09 \\
050713 &    &70 & 9.1e-6 &15-350 &-0.6&  310& R$>$17.7&$>$0.8 & 0.66  & 0.41 \\
050520 &    &80 & 2.4e-6 &20-200 &    &     & R$>$16.1&       &       & 0.02 \\
050504 &    &80 & 1.5e-6 &20-200 &    &     & R$>$16.0&       &       & 0.01 \\
050408 &1.24&34 & 1.9e-6 &30-400 &    &     &Un$>$14.7&$>$5.1 & 0.70  & 0.03 \\
050319 &3.24&10 & 8.0e-7 &15-350 &-1.2&     & R=16.16 &$>$3.4 & 0.48  & 0.01 \\
041219 &    &540& 1.0e-4 &15-200 &    &     & R$>$19.4&$>$1.0 & 1.2   & 1.75 \\
{\bf990123}&1.61&63 & 5.1e-4 &20-1000& 0.2&  720& R=8.95  & 2.0   & 1.65  & 0.02 \\
 \hline 
\end{tabular}
\end{table*}

\begin{figure}
\centerline{\psfig{figure=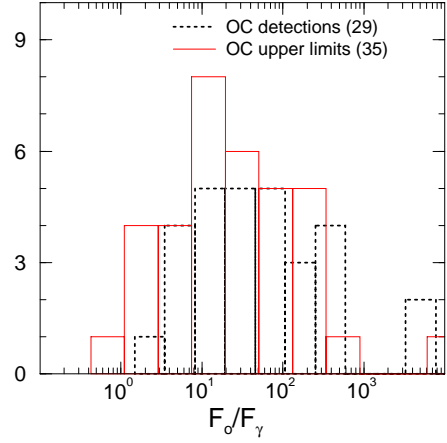,height=6cm}}
\caption{ Distribution of optical counterpart flux to GRB spectral peak flux ratio 
   for bursts with optical counterpart measurements (listed in Table 1). For more
   than half of bursts, only upper limits on the optical counterpart flux have been 
   obtained, which sets an upper limit on the $F_o/F_\gamma$ ratio, leading to a
   lower limit on the Compton parameter of the second scattering and on the GeV emission.
   For the bursts without a determined GRB peak energy $\eg$, we assumed $\eg = 200$ keV, 
   which is the average value for the bursts of Table 1 with measured $\eg$. }
\label{f3}
\end{figure}

  The average optical flux for the 35 upper limits of Table 1 is 1 mJy, while that of 
the 19 counterpart detections 3 mJy, thus upper limits are, on average, 1 mag deeper 
than detections, but both averages have large dispersions (2.8 and 1.6 mag, respectively).
The burst spectral peak energy $\eg$ has been measured for 27 of the GRBs in Table 1,
the average being $\overline{\eg}= 210$ keV with a dispersion of 0.45 dex. For those 
bursts, the flux $F_\gamma$ at the peak, calculated from the GRB fluence and spectrum
(if not known, the low-energy GRB spectral slope was assumed to be $\alpha=0$; the
high-energy spectral slope at $\e > \eg$ was set at $\beta=1.5$), has an average 
$\overline{F_\gamma} = 0.3$ mJy with a dispersion of 0.5 dex. 
To calculate the GeV output for all bursts, we assume the average $\overline{\eg}$
for the 37 burst without a reported peak energy. 
The average optical to GRB peak flux ratio, $F_o/F_\gamma$, which is an important
parameter for the calculation of the GeV emission flux, is about the same for the 
bursts with known $\eg$ ($\overline{F_o/F_\gamma} = 30$) as for those with assumed 
$\eg = \overline{\eg}$ ($\overline{F_o/F_\gamma} = 15$), as can be seen in Figure \ref{f3}.

\subsection{Expected GeV prompt flux}

\begin{figure}
\centerline{\psfig{figure=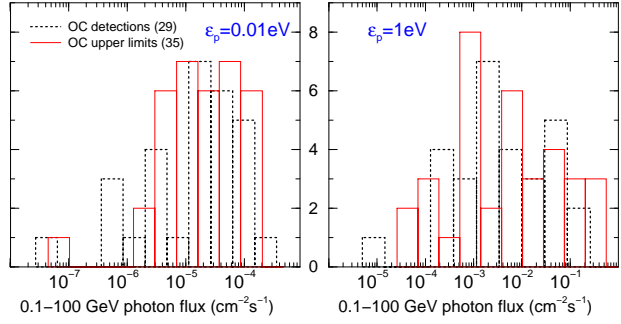,height=4.3cm}}
\caption{ Distribution of the prompt 0.1--100 GeV photon flux resulting from the second 
   scattering of the primary synchrotron emission, for two values of the peak energy 
   $\ep$ of the synchrotron emissivity, and for the optical counterpart and $\gamma$-ray 
   prompt emission properties of the bursts listed in Table 1. Upper limits on the
   optical counterpart flux set lower limits on the GeV flux. 
   For the Fermi-LAT area of $\sim 5000\,{\rm cm^2}$ and a burst lasting for 100 s 
   (the average of the durations given in Table 1), the photon fluxes shown in left 
   panel correspond to 0.2--50 GeV photons collected during the burst, while those in 
   the right panel to $10-10^5$ prompt GeV photons.
   The effect of photon-photon attenuation, which depends strongly on the source Lorentz
   factor and radius where the prompt emission is released, is expected to be negligible 
   for $\ep = 0.01$ eV (left panel), but could completely suppress the GeV photon flux 
   for $\ep = 1$ eV (right panel) if the source radius is less than about $10^{16}$ cm. }
\label{f4}
\end{figure}

 Using the equations of the previous section, we calculate the break energies of the
twice upscattered prompt emission and the peak flux of the GeV spectrum, and integrate
over the power-law piecewise spectrum to obtain the 0.1--100 GeV prompt photon flux 
expected for the bursts of Table 1 in the synchrotron self-Compton model. The distribution 
of the resulting GeV fluxes is shown in Figure \ref{f4} for two values of $\ep$.
For its collecting area of thousands of ${\rm cm^2}$, the LAT onboard the Fermi satellite 
would detect hundreds to tens of thousands of photons during a 100 s burst if the peak 
energy of the synchrotron spectrum were at $\ep = 1$ eV. However, the received GeV photon
flux can be greatly affected by photon-photon attenuation, which depends primarily on the
the source radius $R$. Calculations show that, for the bursts of Table 1, photon-photon 
attenuation is negligible if $R>10^{16}$ cm, but suppresses the 0.1--100 GeV flux above 
$\e_p = 6\times 10^{-3} (F_o/1\,{\rm mJy})^{0.6} (F_\gamma/1\,{\rm mJy})^{-1.5} (R/10^{15}\,
{\rm cm})^2$ eV if $R < 10^{16}$ cm. Thus, the non-detection of a GeV prompt emission 
produced by the second scattering may be due to either (1) an intrinsically weak GeV 
output, when $\ep$ is well below optical, in which case pair-formation is negligible, 
or (2) photon-photon attenuation suppressing the intrinsically-bright GeV emission 
produced when $\ep$ is close to the optical. 

 Fermi-LAT has received more than 10 photons above 1 GeV during GRB 080916C (Tajima et al 
2008) and a similar number of photons below 1 GeV during GRB 080825C (Bouvier et al 2008). 
Optical counterpart measurements are not available for these bursts but the distributions 
shown in Figure \ref{f4} suggest that $\ep$ was in the 0.01--1 eV range, as a burst-integrated 
flux of 10 photons corresponds to $\siml 10^{-4}\, {\rm photons/cm^2 s}$, which is at the 
bright end of the distribution shown for $\ep = 0.01$ eV (left panel) and at the dim end 
for $\ep=1$ eV (right panel). In fact, the above range for $\ep$ set by GeV observations
of 080916C and 080825C is an upper limit because the measured high-energy fluxes are
consistent with the extrapolation of the sub-MeV burst spectrum to GeV, as can be shown
using the burst fluences and high energy spectral slope reported by van der Horst \&
Goldstein (2008). 

 From equation (\ref{PhiGeV}), correlations are expected between the prompt GeV fluence 
$\Phi_{GeV}$ and optical or GRB prompt emission properties (optical flux $F_o$, GRB fluence
$\Phi_\gamma$, burst peak energy $\eg$), provided that the peak energy $\ep$ of the 
synchrotron spectrum has a narrow distribution. 
For the 29 bursts of Table 1 with optical counterpart measurements, we find a significant
correlation only between the expected GeV fluence and the observed sub-MeV fluence if 
$\ep$ is above optical (linear correlation coefficient $r (\log \Phi_{GeV}, \log \Phi_\gamma) 
\simeq 0.75$ corresponding to a $10^{-6}$ probability of a chance correlation), with
other expected correlations being much less significant, owing to the scatter in the
optical and GRB properties and to correlations among them. The strongest such correlation
found is that between the burst fluence and burst peak energy\footnotemark -- $r (\log 
\Phi_\gamma, \log \eg) = 0.70$, with best fit $\Phi_\gamma \propto \eg^{2.0}$w -- which  
weakens the expected $\Phi_{GeV} - \Phi_\gamma$ correlation for $\ep$ is below optical, 
as in this case $\Phi_{GeV} \propto \Phi_\gamma/\eg^3$ (equation \ref{PhiGeV}). 
Other correlations that we found among the prompt optical and GRB properties and which 
have a probability for a chance occurrence less than 10 percent are the optical counterpart 
flux $F_o$ 
(1) correlation with the burst fluence $\Phi_\gamma$, 
(2) anticorrelation with burst duration $t_\gamma$, 
both of which weaken the $\Phi_{GeV} - F_o$ anticorrelation expected from equation 
(\ref{PhiGeV}), and 
(3) correlation with GRB peak energy $\eg$, which strengthens the expected $\Phi_{GeV} -
 F_o$ anticorrelation.
 \footnotetext{ The $\Phi_\gamma - \eg$ correlation was first noticed by Lloyd, Petrosian 
\& Mallozzi (2000), who suggested that it arose from a correlation of intrinsic source
properties (see also Amati et al 2000)}

 To assess the effect of $\ep$ not being universal on the expected correlations, 
we assume that, for the 29 bursts of Table 1 with optical counterpart measurements, 
$\log \ep$ has a uniform distribution between 0.01 eV and 100 eV, and find that such a 
distribution of $\ep$ among bursts weakens the $\Phi_{GeV} - \Phi_\gamma$ correlation
found for a universal $\ep$ above optical, the linear correlation coefficient $r (\log 
\Phi_{GeV}, \log \Phi_\gamma) \in (0.2,0.3)$, corresponding to a 10--30 percent of a 
chance correlation. We conclude that, if the synchrotron self-Compton model for the 
GRB emission is correct, then the measured, prompt GeV fluence (produced by the second 
upscattering) is likely to be correlated with the burst fluence, and less likely to be
correlated with other properties of the prompt emission (e.g. optical counterpart flux), 
but we note that the strength of this conclusion depends on the actual width of the 
synchrotron peak energy distribution.

\subsection{Burst energetics}

 For bursts with known redshift, the isotropic radiative output $\calE_r$
(synchrotron + 1st inverse Compton + 2nd inverse-Compton) can be calculated from the 
total prompt fluence $\Phi = \Phi_{sy} + \Phi_\gamma + \Phi_{GeV} = (Y_\gamma^{-1} + 
1 + Y_{GeV}) \Phi_\gamma$, with the GRB fluence $\Phi_\gamma$ in the 10 keV--10 MeV 
range calculated from the fluences reported in Table 1, using the burst spectrum. 
The resulting  $\calE_r$ for the 34 bursts of Table 1 with known redshift ranges from 
$10^{52}$ to $10^{54}$ erg, with an average of $10^{53.3}$ erg for 20 bursts with 
known peak energy $\eg$ and $10^{53.0}$ erg for all 34 bursts, assuming $\eg = 200$ 
keV when not known.

\begin{figure}
\centerline{\psfig{figure=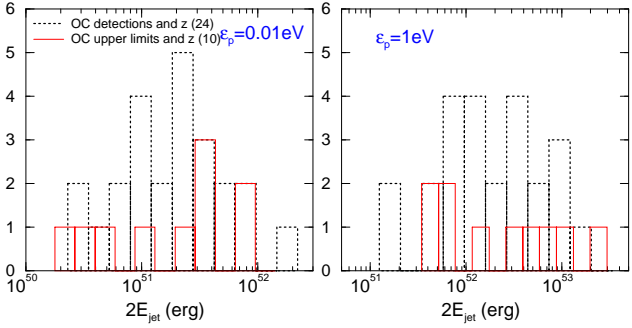,height=4.4cm}}
\caption{ Distribution of lower limits on the collimation-corrected outflow initial energy 
  (assuming a two-sided jet), for the optical counterparts and GRB prompt emission properties 
  of Table 1, and for two possible values of the peak energy $\ep$ of the synchrotron 
  emissivity. A radiative efficiency ($\eta$) for the prompt emission of 50 percent was 
  assumed. The lower limit on the jet opening was determined from the latest epoch $t_b$ 
  (Table 1) until which the optical afterglow light-curve decay does not exhibit the 
  steepening expected from "seeing" the jet boundary, and assuming that the ambient medium 
  has the typical density expected for a Wolf-Rayet progenitor of long bursts. 
  Whenever it cannot be determined through afterglow observations, $t_b = 1$ day was assumed.
  (The jet energy has a moderate dependence on these parameters: $E_j\propto (t_b/\eta)^{1/2}$.) 
  Only the bursts of Table 1 with known redshift have been used and $\eg = 200$ keV was assumed 
  when not known. 
  Compared to the $\ep = 1$ eV case (right panel), the required jet energy is found to
  decrease by a factor $\sim 10$ for a factor 100 increase or decrease in $\ep$,
  as illustrated in the left panel for $\ep =0.01$ eV. } 
\label{f5}
\end{figure}

 The true radiative output of GRBs depends on the degree of ejecta collimation.
If the optical light-curve breaks (i.e. decay steepenings) observed at 0.3--3 days
in a majority of well-monitored pre-Swift bursts (e.g. Zeh, Klose \& Kann 2006) are 
due to "jet effects" (i.e. boundary of the jet becoming visible to observer when
its decreasing Lorentz factor $\Gamma$ reaches $\theta_j^{-1}$, the inverse of the
jet half-opening angle, and jet lateral spreading beginning to affect the jet dynamics
at about the same time), then the epoch $t_b$ of the light-curve break can be used to
determine the jet opening $\theta_j$:
\begin{equation}
 \theta_j = [\Gamma(t_b)]^{-1} = 0.096\, \left[\frac{t_{b,d}}
            {(z+1)\calE_{k,53}}\right]^{1/4} \; {\rm rad}
\end{equation}
where $t_b$ is measured in days and $\calE_{k,53}$ is the isotropic-equivalent of the
jet kinetic energy after the prompt phase, measured in $10^{53}$ erg. The derivation
of the above result for the jet dynamics $\Gamma(t)$ assumed that the jet is decelerated
by its interaction with the wind produced by a Wolf-Rayet star. The post-burst jet
kinetic energy is not known but can be related to the GRB bolometric output $\calE_r$
by assuming a burst radiative efficiency $\eta = \calE_r/(\calE_r + \calE_k)$. 
Then the collimation-corrected initial energy of the two-sided GRB jet is
\begin{equation}
 2E_k = \frac{1}{2} \theta_j^2 (\calE_r + \calE_k) = 10^{50.7} 
     \left[ \frac{\calE_{r,53}\, t_{b,d}}{(z+1)\eta(1-\eta)} \right]^{1/2} {\rm erg} \;.
\label{Ejet}
\end{equation}
Monitoring of the optical emission of the GRB afterglows listed in Table 1 is somewhat
limited, nearly none of the optical light-curves displaying a jet-break until the last
measurement, as shown by slow optical decays $d\log F_\nu /d\log t$ listed in column 10 
of Table 1 (decays faster than $t^{-2}$ are likely to be caused by a jet-break having 
occurred). Evidence for jet-breaks in the X-ray afterglow light-curve, consisting of
a steepening to a decay faster than $t^{-2}$, is not considered here because the decoupled
optical and X-ray afterglow light-curve behaviours seen in many cases (e.g. chromatic
X-ray light-curve breaks) suggests that sometimes these two emission arise from different
mechanisms and/or parts of the relativistic outflow. Thus, for most afterglows, we have 
only a lower limit on $t_b$ (column 9 in Table 1) which yields a lower limit on the initial 
jet energy $E_k$. 

 Figure \ref{f5} shows the distribution of the lower limits on $E_k$ for bursts with
known redshift, assuming a GRB bolometric radiative efficiency $\eta=0.5$ (which minimizes
the jet energy -- equation \ref{Ejet}) and $t_j = 1$ day whenever the available optical 
afterglow monitoring does not allow us to set even a lower limit on $t_b$. 
For a given burst, the jet energy is maximal for $\ep$ in the optical, as this value
minimizes the synchrotron peak flux required to account for the observed optical 
counterpart flux, which maximizes the Compton parameter, the GeV output, and the 
total isotropic-equivalent burst output.
The largest lower limit on the initial jet kinetic energy, obtained for $\ep =1$ eV, is 
$10^{53}$ erg, being lower by a factor 10 for a factor 100 increase or decrease in $\ep$.  

 The energy that the long-GRB progenitor (black-hole plus accretion torus formed after 
the collapse of a Wolf-Rayet core) can deposit into a relativistic jet depends on the 
available energy reservoir (a torus with a rest-mass energy $1\, M_\odot c^2 = 2\times 
10^{54}$ erg and a black-hole with a comparable spin energy, for the collapsar model -- 
Woosley 1993) and the efficiency at which the available energy is extracted and 
deposited into highly relativistic ejecta [e.g. magnetohydrodynamical energy extraction 
is limited to 5.7 percent (for a non-rotating BH) and 42 percent (for a maximally rotating 
BH) of the torus gravitational binding energy and up to 29 percent of the black-hole mass].
 In the case when the accretion rate is $0.1\, M_\odot {\rm s^{-1}}$ and the black-hole 
spin parameter is $a=0.95$, Popham, Woosley \& Fryer (1999) obtain a maximum of $10^{52.3}$ 
erg for the energy of the outflow resulting from the annihilation of neutrinos and 
antineutrinos produced by dissipation in the torus. 
 Using general relativistic MHD simulations of accreting tori by Kerr black-holes, Krolik, 
Hawley \& Hirose (2005) find that, at the radius of marginal stability, the Poynting flux 
produced by the Blandford-Znajek mechanism carries 0.25 percent of the accreted mass-energy
for $a=0.5$, 1 percent for $a=0.9$, rising to about $\sim 10$ percent for $a=0.998$. 
Similar results, showing a rapidly increasing jet efficiency as function of the BH spin
parameter, are obtained by McKinney (2005), who obtains an upper limit of 6.8 percent 
for the jet efficiency for a maximally rotating BH, corresponding to a jet energy of 
$10^{53.1}$ erg, if the disk as a mass of $1\,M_\odot$.

 Thus, the jet energy expected in the collapsar model is less than $10^{53}$ erg,
which implies that, if the sub-MeV emission of the GRBs listed in Table 1 was produced 
from upscattering of a lower energy emission, the peak energy $\ep$ of the primary spectrum 
could not have been in the optical for all bursts (right panel of Figure \ref{f5}). 
Instead, if $\ep$ were universal, the synchrotron peak energy must have been below 
0.1 eV (left panel of Figure \ref{f5}) or above 10 eV, to yield lower limits on the
jet energy that are sufficiently below the theoretical upper limit of $10^{53}$ erg.
Values of $\ep$ above 10 eV lead to even lower required jet energies but the 10 keV prompt 
flux could be dominated by the synchrotron emission instead of the first upscattering, thus 
the resulting low-energy GRB spectrum would be softer than usually observed. For this 
reason, we consider that only $\ep < 0.1$ eV is a possible solution for reducing the 
required jet energetics below the theoretical expectation for the collapsar model.

\section{Conclusions}

 The occasional detection of an optical counterpart whose brightness exceeds the
extrapolation of the GRB sub-MeV emission suggests that, if the two emissions arise
from same medium, the burst could be the first upscattering of the synchrotron spectrum 
that yields the optical counterpart. We find 10 such cases in a sample of 29 bursts with 
optical counterpart measurements, but the true fraction of over-bright optical counterparts 
could be larger because half of those 29 bursts have been observed only by Swift-BAT
(Table 1), which underestimates the true hardness of the GRB spectral slope below the 
peak energy if that peak energy fell in BAT's relatively narrow bandpass.

 A straightforward expectation for the synchrotron self-Compton model for GRBs is that
the upscattering of the burst photons yields a GeV--TeV prompt emission, whose brightness
is found to depend strongly on the peak energy of the synchrotron spectrum. 
In this paper, we provided the formalism by which the GeV prompt emission from the second
scattering is related to the sub-MeV emission from the first scattering and the optical
emission from synchrotron, and applied that formalism to bursts with optical counterpart
measurements, to estimate the expected GeV prompt fluxes, the bolometric GRB output and
energetics, and the correlation of the expected GeV fluence with the burst and optical
brightnesses.

 The synchrotron self-Compton model can be tested in the following ways. The measurement 
by Fermi-LAT or Agile-GRID of the 0.1--100 GeV fluence of the emission produced during 
the burst by the second upscattering, combined with the properties (peak flux and energy) 
of the prompt burst emission produced by the first upscattering, leads to two solutions 
for the location of the synchrotron peak energy (equation \ref{PhiGeV}). The lower energy
solution corresponds to a soft optical spectrum and a hard GeV spectrum, while the higher 
energy solution is identified with a hard, self-absorbed optical and a soft (falling) 
GeV spectrum. Future multicolour measurements of the optical prompt emission obtained 
by fast-response telescopes and measurements of the GeV emission by high-energy 
satellites will allow a test of consistency between the observed optical and GeV spectra 
and the above-mentioned model expectations. A stronger test will be possible if the peak 
energy of the second scattering emission spectrum is also determined, as the GeV flux and
peak energy provide two independent determinations of the synchrotron spectral peak, 
for given properties of the $\gamma$-ray spectrum (see Figure \ref{f2}). 

 From the expected GeV output and using the constraints on the outflow opening set by 
the afterglow optical light-curve, we have calculated lower limits on the collimated 
radiation output and jet initial energy for 34 bursts with optical counterpart measurements 
and redshifts. The resulting jet energies are lower limits because one third of optical
counterpart measurements are upper limits, which lead to lower limits on the GeV output,
and because most of the available coverage of optical afterglows sets only lower limits 
on the jet-break time, leading to lower limits on the jet half-angle. 
Figure \ref{f5} shows that the resulting lower limits on the double-jet initial energy 
ranges over 2 decades, with the largest value ($10^{53}$ erg) being obtained if the peak 
energy of the synchrotron spectrum is close to the optical (right panel), and with the 
lower limit on the jet energy decreasing by a factor 10 for a factor 100 decrease 
of the synchrotron peak energy (left panel).
Thus, the energetics required by the synchrotron self-Compton model for GRBs and the 
upper limit of $\sim 10^{53}$ erg expected for jets produced after the core-collapse 
of massive stars indicate that the peak energy of the synchrotron spectrum should be 
often well below optical. 

 For the 29 bursts with optical counterparts measurements, we find that, if the unknown
peak energy of the synchrotron spectrum does not have a very wide distribution, the 
brightness of the second inverse-Compton scattering remains {\sl correlated} with the flux 
of the first upscattering and {\sl anticorrelated} with that of the primary synchrotron 
spectrum. Thus, the synchrotron self-Compton model for GRBs will be {\sl invalidated} 
if Fermi-LAT detects GeV prompt emission consistent with the extrapolation of the burst 
spectrum for bursts that are bright at sub-MeV energies and dim in the optical. This 
test of the synchrotron self-Compton model for GRBs applies only if GeV prompt photons 
are detected because the lack of such detections in a given burst may not necessarily 
imply the production of low GeV prompt fluxes, but could be due instead to the source
being optically-thick to photon-photon attenuation.

\section*{Acknowledgments}
 The author acknowledges the great help in collecting the optical counterparts measurements
provided by the {\sl GRBlog} site at {\sl http://grad40.as.utexas.edu} created by Robert 
Quimby and maintained together with Erin McMahon and Jeremy Murphy.  \\
This work was supported by the US Department of Energy through the LANL/LDRD 20080039DR 
program.


\begin{thebibliography}{99}
\bibitem{} Amati L. et al, 2002, A\&A, 390, 81
\bibitem{} Asano K., Inoue S., 2007, ApJ, 671, 645
\bibitem{} Bouvier A. et al, 2008, GCN 8183
\bibitem{} Galama T. et al, 1999, Nature, 398, 394
\bibitem{} Gonzales M. et al, 2003, 424, 749
\bibitem{} Granot J., Guetta D., 2003, ApJ, 598, L11           
\bibitem{} van der Horst A., Goldstein A., 2008, GCN 8141, 8278
\bibitem{} Hurley K. et al, 1994, Nature, 372, 652
\bibitem{} Kaneko Y. et al, 2008, ApJ, 677, 1168
\bibitem{} Karpov S. et al, 2008, GCN 7558
\bibitem{} Krolik J., Hawley J., Hirose S., 2005, ApJ, 622, 1008
\bibitem{} Kumar P., Panaitescu A., 2008, preprint (arXiv:0805.0144)
\bibitem{} Lloyd N., Petrosian V., Mallozzi R., 2000, ApJ, 534, 227
\bibitem{} M\'esz\'aros P.. Rees M., 1997, ApJ, 476, 232 
\bibitem{} M\'esz\'aros P.. Rees M., 1999, MNRAS, 306, L39
\bibitem{} McKinney J., 2005, ApJ, 630, L5
\bibitem{} Panaitescu A., M\'esz\'aros P., 2000, ApJ, 544, L17
\bibitem{} Panaitescu A., Kumar P., 2007, MNRAS, 376, 1065
\bibitem{} Papathanassiou H., M\'esz\'aros P., 1996, ApJ, 471, L91
\bibitem{} Pe'er A., Waxman E., 2004, ApJ, 603, L1 
\bibitem{} Perley D. et al, 2008, ApJ, 672, 449
\bibitem{} Piran T., Sari R., Zou Y., 2008, preprint (arXiv:0807.3954)
\bibitem{} Popham R., Woosley S., Fryer C., 1999, ApJ, 518, 356
\bibitem{} Racusin J. et al, 2008, Nature, 455, 183
\bibitem{} Rees M., M\'esz\'aros P., 1994, ApJ, 430, L93
\bibitem{} Sari R., Piran T., 1999, ApJ, 517, L109
\bibitem{} Sommer M. et al, 1994, ApJ, 422, L63
\bibitem{} Stamatikos M. et al, 2008, preprint (arXiv:0809.2132)
\bibitem{} Tajima H. et al, 2008, GCN 8246
\bibitem{} Wang X., Dai Z., Lu T., 2001, ApJ, 556, 1010 
\bibitem{} Woosley S., 1993, ApJ, 405, 273
\bibitem{} Yu Y., Wang X., Dai Z., 2008, preprint (arXiv:0806.2010)
\bibitem{} Zeh A., Klose S., Kann D., 2006, ApJ, 637, 889
\bibitem{} 
\end{thebibliography}
\end{document}